\documentclass{emulateapj}
\usepackage{apjfonts}
\usepackage{natbib}
\usepackage{graphicx}
\usepackage{amssymb,amsmath}
\usepackage{xcolor}
\usepackage[export]{adjustbox}
\usepackage[citebordercolor=green]{hyperref}
\shorttitle{Gibbs Sampling}
\usepackage{array}
\newcolumntype{P}[1]{>{\centering\arraybackslash}p{#1}}
\usepackage{hyperref}

\begin{document}

\title{A Partial-Sky Gibbs ILC Approach for the Estimation of CMB Posterior over Large Angular Scales of the Sky}
\author{Vipin Sudevan\altaffilmark{1}, Ujjal Purkayastha\altaffilmark{2} \& Rajib Saha\altaffilmark{2}}
\altaffiltext{1}{LeCosPA, National Taiwan University, Taipei, Taiwan 10617}
\altaffiltext{2}{Department of Physics, Indian Institute of Science
Education and Research,  Bhopal, M.P, 462023, India.}

\begin{abstract}
{In this article we present a formalism to incorporate the partial-sky maps to the Gibbs ILC 
algorithm to estimate the joint posterior density of the Cosmic Microwave Background (CMB) signal and 
the theoretical CMB angular power spectrum given the 
observed CMB maps. In order to generate the partial-sky maps we mask all the observed 
CMB maps provided by WMAP and Planck satellite mission using a Gaussian smoothed mask formed on the basis of 
thermal dust emissions in Planck 353 GHz map. The central galactic region from all the input maps is 
removed after the application of the smoothed mask.  
While implementing the Gibbs ILC method on the partial-sky maps, we convert the partial-sky 
cleaned angular power spectrum to the full-sky angular power spectrum using the mode-mode 
coupling matrix estimated from the smoothed mask.  The main products of our analysis are partial-sky cleaned best-fit CMB 
map and an estimate of the underlying full-sky 
theoretical CMB angular power spectrum along with their error estimates. We validate the 
methodology by performing detailed Monte Carlo simulations after using realistic models of  
foregrounds and detector noise consistent with the WMAP and Planck frequency channels used in our 
analysis. We can estimate the posterior density and full-sky theoretical CMB 
angular power spectrum, without any need to explicitly model 
the foreground components, from partial-sky maps using our method.  Another important feature of this method 
is that the power spectrum results along with the error estimates 
can be directly used for cosmological parameter estimations. }
\end{abstract}


\section{Introduction}
\label{into}
Accurate measurement of the fluctuations present in the temperature and polarization fields of CMB anisotropies 
provides us with a wealth of knowledge regarding the origin, geometry and composition  of our  Universe  (e.g.,  see~\cite{Aghanim:2018eyx}).  
However, strong emissions in the microwave 
region of the frequency spectrum by various astrophysical sources, the foregrounds,  
hinders us from directly observing the underlying true 
CMB  signal. Therefore, in order to disinter the physics encoded in the 
CMB anisotropies the foremost challenge is to properly remove the foregrounds 
present in the CMB observations and thereby to recover the underlying CMB signal.

One of the important methods to minimize the foregrounds in a (foreground) model-independent 
manner is the so-called the {\it internal-linear-combination} (ILC) 
method~\citep{Tegmark96, Tegmark2003, Bennett2003, Eriksen2004, Hinshaw_07, Saha:2005aq, Saha:2007gf, Saha2011,Saha:2015xya, Sudevan2017}. 
A unique feature of the ILC method is that it does not require an explicit modeling of 
foregrounds in the form of templates or the spectral index information in order to remove them. 
This method purely relies on the assumption that the CMB photons follow a blackbody 
distribution along every direction of the sky so that its temperature fluctuations 
 in thermodynamic units is independent of frequency. 
In the ILC method, foreground minimization is achieved by linearly superposing all the available 
foreground contaminated observed CMB maps with some amplitude terms, the weight factors.  
These weights can be  computed analytically by minimizing the variance of the 
cleaned map. Instead of using the usual unweighted variance, some of the authors developed a new method, 
the global ILC method~\citep{Saha2017},  where the weights are estimated 
by minimizing a theoretical CMB covariance weighted variance. 
This method outperforms the usual ILC over large angular scales of the sky. Unlike the usual ILC, 
the new global ILC weights could effectively 
minimize the contributions from chance-correlations between CMB and 
foregrounds over large angular scales of the sky.  Some of the authors, ~\cite{Sudevan:2018qyj, Sudevan:2020tvl},  
developed a Gibbs ILC method to estimate 
the CMB posterior density and corresponding  theoretical CMB angular power spectrum given the observed data over 
the large angular scales of the sky in a (foreground) model-independent manner.  The Gibbs ILC method involves two important 
steps:
\begin{itemize}
\item Given the observed CMB data and thoretical CMB angular power spectrum, 
a cleaned CMB map is sampled by minimizing the foregrounds present in the 
observed CMB maps using the global ILC method.
\item Sample a new theoretical CMB angular power spectrum from its conditional density 
given the cleaned CMB map and observed CMB data.
\end{itemize} 

To implement  Gibbs ILC method in pixel space is a computationally exhaustive task since it involves 
estimation of cleaned CMB maps using global ILC method (as discussed in~\cite{Saha2017}) during 
each Gibbs iteration.  
In~\cite{Sudevan:2018qyj}, some of the authors implemented global ILC method in  spherical harmonic 
 basis in order to preform foreground removal.
Further, the efficiency of the foreground removal 
using the global ILC method improves when one carries out 
the foreground  minimization in an iterative manner~\citep{Saha2017}. Now taking into account that the  Gibbs ILC method 
discussed in~\cite{Sudevan:2018qyj, Sudevan:2020tvl, Purkayastha:2020amx, Yadav:2020elw} 
uses full-sky CMB observations, it is, therefore, only natural to ask whether the Gibbs ILC method can be generalized such that  
it can handle maps with incomplete-sky data?

 The challenges of  dealing with incomplete-sky maps are the following:
 \begin{itemize}
 \item Given partial-sky observed CMB maps and partial-sky cleaned CMB map, how to sample a theoretical CMB 
 angular power spectrum?
 \item How to incorporate a full-sky CMB angular power spectrum into the global ILC algorithm during a partial-sky analysis?
{ \item Finally,  how to address the issue of ``Gibbs phenomenon''~\footnote{Refers to the peculiar manner 
 in which the Fourier representation behaves at a jump discontinuity~\cite{1898Natur..59..200G, medical_imaging, signal_analysis, Gibbs_phenomenon}.} in a spherical harmonic analysis  involving discrete masks (i.e., mask with pixel values 1 and 0 only)?.}   
 \end{itemize}
{ In order to} develop a truly partial-sky Gibbs ILC method these three questions need to be answered.  

In this article, we incorporate the MASTER approach~\citep{master} to 
our Gibbs ILC algorithm to sample a partial-sky CMB angular spectrum given the full-sky spectrum. 
As for Gibbs phenomenon, there exist atleast 
two possible solutions 
for suppressing its  adverse affects. One of them  is to allow the   spherical 
harmonic expansion to infinitely high multipoles~\citep{medical_imaging, signal_analysis, Gibbs_phenomenon}.  
Another option is to use a mask smoothed by a Gaussian beam to minimize the jump discontinuities at 
the region boundaries. Since there exists 
a maximum multipoles $\ell_{max}$ upto which spherical harmonic coefficients can be expanded for a given 
(finite) pixel resolution of the HEALPix~\footnote{{\bf H}ierarchical  {\bf E}qual  {\bf A}rea  Iso-{\bf L}atitude  {\bf Pix}elation  
of  Sphere  which  was developed by~\cite{Gorski2005}.} map, utilizing a smoothed mask is, therefore,  a more viable option. 
Hence a work-around to suppress 
the Gibbs phenomenon effectively in our Gibbs ILC method is to use full-sky observed CMB data multiplied with a  smoothed mask. 

We organize the paper as follows. We discuss in Section~\ref{formalism} the basic formalism of our approach.  
We describe the  input maps, masks that we use in this analysis in Section~\ref{input}.  In  Section~\ref{methodology} we 
lay out our procedure and discuss how to estimate the posterior density in a partial-sky analysis. We  present  our  results  
of  analysis  of  WMAP  and  Planck frequency maps at low resolution in Section~\ref{Results}. 
In Section~\ref{MC} we  validate our method 
for estimating the joint posterior density of CMB signal and its theoretical angular power spectrum by performing  detailed 
Monte Carlo simulations. Finally, we conclude in Section~\ref{Conclusion}.

\section{Formalism}
\label{formalism}
\subsection{The Gibbs ILC Method}  
\label{GibbsILC}
In the Gibbs ILC method, to estimate the joint CMB posterior density, $P({\bf S}, C_\ell \vert {\bf D})$, 
 we draw samples of ${\bf S}$ and $C_\ell$ from their respective conditional densities using the Gibbs sampling~\citep{Eriksen2008, Eriksen2007, Gibbs1984} technique. Here 
 ${\bf S}$, $C_\ell$ and ${\bf D}$ represent the true CMB signal, theoretical CMB angular power spectrum, and 
the observed CMB data respectively. At the 
beginning of some iteration $(i+1)$ in a Gibbs sampling procedure, a cleaned CMB signal ${\bf S}^{i+1}$ and a theoretical 
CMB angular power spectrum $C_\ell^{i+1}$ is sampled from their respective conditional densities 
$P_1({\bf S} \vert {\bf D}, C_\ell)$ and $P_2(C_\ell \vert {\bf D}, {\bf S})$ as follows,
\begin{equation}
{\bf S}^{i+1}  \leftarrow P_1({\bf S} \vert {\bf D}, C_\ell^i)\, , \\
\label{sample1}
\end{equation}
\begin{equation}
C_\ell^{i+1} \leftarrow  P_2(C_\ell \vert {\bf D}, {\bf S}^i) \, .
\label{sample2}
\end{equation}
Using the pair of samples, ${\bf S}^{i+1}$ and $C_\ell^{i+1}$ generated at the end of $(i+1)^{\tt th}$ iteration, we repeat 
the above two sampling steps (Eqns.~\ref{sample1} and~\ref{sample2}) for a large number of iterations till convergence is 
achieved. Ignoring few samples generated during the initial (burn-in) phase, rest of the samples of {\bf S} and $C_\ell$ appear as if 
they are sampled from the joint posterior density $P({\bf S}, C_\ell \vert {\bf D})$.

In the Gibbs ILC method, in order to estimate $P({\bf S}, C_\ell \vert {\bf D})$ in a (foreground) model-independent 
manner, we minimize the foregrounds present in the observed CMB maps using the global 
ILC algorithm at each Gibbs step. Using the global ILC method, with CMB maps ${\bf d}_i$ observed at $n$ different frequencies, 
an estimate ${\bf \hat S}$ of the underlying true {\bf S} is obtained as  
\begin{equation}
{\bf \hat S} = \sum_{i=1}^{n} w_i {\bf d}_i = \sum_{i=1}^{n} w_i \sum_{\ell > 0}\sum_{m=-\ell}^{\ell} a_{\ell m}^i Y_{\ell m}({\bf \hat{n}})\, ,
\end{equation}
where $w_i$ is the weight corresponding to the map from $i^{\tt th}$ frequency channel. These weights are subject to 
a constraint that they should sum to unity, i.e., $\sum_{i=1}^{n} w_i = 1$. This constrain ensures that the CMB signal will 
not undergo any undesired modification during the entire foreground removal procedure. 
By maximizing the likelihood of the model given the full-sky observed CMB data and the theoretical CMB angular 
power spectrum in the spherical-harmonic domain, the  conditional density $P_1({\bf S} \vert {\bf D}, C_\ell)$ 
is obtained as follows:  
\begin{eqnarray}
P_1({\bf S} \vert {\bf D}, C_\ell)  \propto \prod_{\ell, m} e^{-\sum_{i,j}w_{i}a_{\ell m}^{i}a_{\ell m}^{\ast j}w_j / C_{\ell}}  \\ \propto   e^{-\sum_{i,j}w_{i}w_j {\sum_{\ell} (2\ell + 1) \hat{\sigma}_{\ell}^{ij} / C_\ell} } \, .
\label{hilc_new}
\end{eqnarray} 
${\hat{\sigma}_\ell^{ij}}$ in the above equation is the cross power spectrum between the observed CMB maps ${\bf d}_i$ 
and ${\bf d}_j$. Form the Eqn.~\ref{hilc_new}, we define an estimator $\sigma^{2}_{r}$ given by 
\begin{equation}
{\sigma}_{r}^2 = \sum_{i,j}w_{i}w_j  \sum_{\ell=2}^{\ell_{\tt max}} (2\ell + 1) \frac{\hat{\sigma}_\ell^{ij}}{{C}_\ell^{\prime}} \, ,
\label{hilc}
\end{equation}
which is then minimized in order to 
obtain the weights. ${C_\ell^{\prime}}$ (= $C_\ell B_{\ell}^2 P_{\ell}^2$) in Eqn.~\ref{hilc} is the 
beam and pixel smoothed theoretical CMB angular power spectrum ($C_\ell$).    
The choice of weights which 
minimizes $\sigma_{r}^2$ is obtained by following a Lagrange's multiplier approach,
\begin{equation}
{\bf W} = \frac{{\bf {\hat A}^\dagger} {\bf e}}{{\bf e}^T {\bf {\hat A}^\dagger} {\bf e}}\, ,
\label{weight}
\end{equation}
where, 
\begin{equation}
{\hat A}_{ij} = \sum_{\ell=2}^{\ell_{\tt max}} (2\ell + 1) \frac{\hat{\sigma}_\ell^{ij}}{C_\ell^{\prime}} \, . 
\label{hilc1}
\end{equation}
{\bf W} is a $(n \times 1)$ weight vector and {\bf e} is a $(n \times 1)$ shape vector of CMB in thermodynamic units.

To draw the samples of $C_\ell$ given ${\bf S}$ and {\bf D} in the Gibbs ILC method, we first 
define a variable $z = {\hat C}_\ell (2\ell + 1) / C_\ell$, where ${\hat C}_\ell$ is estimated from the 
cleaned CMB map. We obtain the conditional density 
$P_2(C_\ell \vert {\bf D}, {\bf S})$  as 
\begin{equation}
P_2(C_\ell \vert {\hat C}_\ell) \propto z^{-(2\ell - 1)/2 - 1} {\tt exp}\big[-\frac{z}{2} \big]\, .
\label{z-dist}
\end{equation}
From Eqn.~\ref{z-dist}, one can infer that the variable 
$z$ follows a $\chi^2$ distribution with $2\ell - 1$ degrees of freedom. 
We  then sample a theoretical CMB angular power spectrum as follows
\begin{equation}
C_\ell = {\hat C}_\ell (2\ell + 1) /z \, ,
\label{cl_sample}
\end{equation}
where $z$ is drawn from the $\chi^2$ distribution of $2\ell - 1$ 
degrees of freedom.

\subsection{Partial-Sky Analysis}
The Gibbs ILC method~\citep{Sudevan:2018qyj, Sudevan:2020tvl}, discussed in Section~\ref{GibbsILC}, 
uses full-sky observed CMB maps provided by WMAP~\citep{2011ApJS..192...14J} and Planck~\citep{Akrami:2018vks} 
CMB observations. 
{ Oftentimes in the point of view of the foreground removal problem it is advantageous to mask certain 
highly contaminated regions of the sky like the Galactic plane, strong point sources etc. These  
masked-observed CMB maps are then fed into the foreground minimization pipeline}. The 
advantage of implementing this approach is that the ILC (and the global ILC)  weights 
can be tuned in such a way that they are estimated based on the strength of the 
foreground contaminations across the sky. A similar  
partial-sky approach is not direct in the case of Gibbs ILC method. Even though one can sample 
a partial-sky CMB map by removing the foregrounds using global ILC method given the partial-sky 
observed maps in pixel space.  However, a similar implementation of the global ILC method in spherical harmonic 
space for the partial-sky maps is not straightforward. Another challenge is to understand how to sample a 
theoretical CMB angular power spectrum 
given a partial-sky cleaned CMB map and partial-sky CMB data.

 In the current analysis, we formalize a technique such that the 
Gibbs ILC method can incorporate only certain regions of the sky (preferably low foreground contaminated regions) from the 
full-sky observations to estimate the joint CMB posterior density and full-sky cleaned theoretical CMB angular 
power spectrum.

\begin{figure*}[!t]
\centering
\includegraphics[scale=0.29]{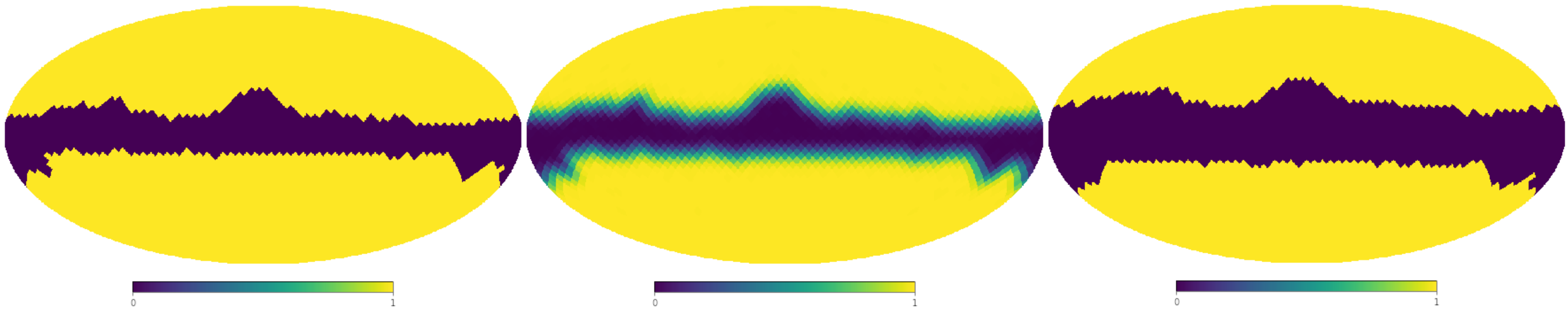}
\caption{We present various masks that is used in this analysis. Left panel shows the mask based on the thermal dust emission, the 
{\tt ThDust5000} mask. The central galactic region of the {\tt ThDust5000} mask (shown in violet) corresponds to very intense thermal dust 
emissions. In the middle panel we show the mask obtained after smoothing the {\tt ThDust5000} mask, the {\tt ThDust5000$\_$Sm},  
using a Gaussian beam of FWHM = $9^\circ$. In the last panel we show the  
effective mask, i.e., {\tt ThDust5000$\_$Eff}, which consists of all pixels set to 1 whose corresponding value in 
{\tt ThDust5000$\_$Sm} mask is $\ge$ 0.95 and rest 0.}
\label{fig_ourmask}
\end{figure*}

\subsubsection{Sampling a Theoretical CMB angular power spectrum}
In this section, we briefly review the MASTER algorithm~\citep{master} implemented in our framework to sample a theoretical 
CMB angular power spectrum from 
its conditional density given a partial-sky cleaned CMB map and partial-sky CMB data. We begin with a 
full-sky CMB anisotropic map which can be expanded in terms of spherical harmonics as follows
\begin{equation}
{\bf S}({\bf \hat{n}}) = \sum_{\ell > 0}\sum_{m=-\ell}^{\ell} a_{\ell m} Y_{\ell m}({\bf \hat{n}}) \, ,
\label{temp}
\end{equation}
with $^{}a_{\ell m} = \int d\Omega\, {\bf S}({\bf \hat{n}})\, Y^{\ast}_{\ell m}({\bf \hat{n}})$ and 
angular power spectrum defined as, 
\begin{equation}
C_{\ell} = \frac{1}{2\ell + 1} \sum_{m=-\ell}^{\ell} \vert a_{\ell m} \vert^2 \, .
\label{cldef}
\end{equation}
A partial-sky map (${\bf \tilde{S}}$) used  
in this analysis is a full-sky map multiplied with a Gaussian smoothed mask ({\bf M}) i.e.,
\begin{equation}
\tilde{\bf S}({\bf \hat{n}}) = {\bf M}({\bf \hat{n}})  {\bf S}({\bf \hat{n}}) \, .  
\label{maskmap}
\end{equation} 
Details about the smoothed mask is given in the Section~\ref{ourmask}. 
Substituting Eqn.~\ref{temp} in Eqn.~\ref{maskmap}, we can expand the masked map in spherical harmonic 
representation as follows 
\begin{eqnarray}
\tilde{a}_{\ell m} &=& \int d\Omega \;  \tilde{\bf S}({\bf \hat{n}}) \; Y^{\ast}_{\ell m} ({\bf \hat{n}}) \\
&=& \sum_{\ell^{'}m^{'}} a_{\ell^{'}m^{'}} K_{\ell m \ell^{'}m^{'}}[M] \, ,
\label{partalm}
\end{eqnarray}	
where $K_{\ell m \ell^{'}m^{'}}$ is the coupling kernel defined as   
\begin{equation}
K_{\ell m \ell^{'}m^{'}} = \int d\Omega\;{\bf M}({\bf \hat{n}})\; Y_{\ell m}({\bf \hat{n}})\; Y_{\ell^{'}m^{'}}^{\ast}({\bf \hat{n}}) \, .
\label{kernel}
\end{equation}

The mask can be expanded in spherical harmonics with the coefficients 
\begin{equation}
a_{\ell m}^{M} = \int  d\Omega \; {\bf M}(\hat{\bf n}) \; Y^{\ast}_{\ell m}(\hat{\bf n}) \, ,
\label{maskalm}
\end{equation}
Now substituting Eqn.~\ref{maskalm} in Eqn.~\ref{kernel}, the coupling kernel reads as 
\begin{equation}
K_{\ell m \ell^{'}m^{'}} = \sum_{\ell^{\prime\prime} m^{\prime\prime}} a_{\ell^{\prime\prime} m^{\prime\prime}}^M \int d\Omega Y_{\ell^{\prime\prime} m^{\prime\prime}}({\bf \hat{n}}) Y_{\ell^\prime m^\prime}^{\ast}({\bf \hat{n}})Y_{\ell m}({\bf \hat{n}})\,  
\label{kernelexp}
\end{equation}
\begin{equation}
= \sum_{\ell^{\prime\prime} m^{\prime\prime}} a_{\ell^{\prime\prime} m^{\prime\prime}}^{M} (-1)^{m^{\prime}} \Big[ \frac{(2\ell^{\prime\prime} +1)(2\ell^{\prime} + 1)(2\ell+1)}{4\pi} \Big]^{1/2}
\end{equation}
\begin{equation*}
\begin{pmatrix}
\ell & \ell^{\prime} & \ell^{\prime\prime} \\ 0 & 0 & 0
\end{pmatrix}
\begin{pmatrix}
\ell & \ell^{\prime} & \ell^{\prime\prime} \\ m & -m^{\prime} & m^{\prime\prime}
\end{pmatrix} \, .
\end{equation*}
$\begin{pmatrix}
\ell & \ell^{\prime} & \ell^{\prime\prime} \\ m & m^{\prime} & m^{\prime\prime}
\end{pmatrix}$ is Wigner 3-$j$ symbol also known as Clebsch-Gordan coefficient which describes 
the coupling of the three angular momentum vectors. After using the orthogonality condition of Wigner Symbol, 
\begin{equation}
\sum_{\ell^{\prime\prime} m^{\prime\prime}} (2\ell^{\prime\prime} + 1) \begin{pmatrix}
\ell & \ell^{\prime} & \ell^{\prime\prime} \\ m & m^{\prime} & m^{\prime\prime}
\end{pmatrix}
\begin{pmatrix}
\ell & \ell^{\prime} & \ell^{\prime\prime} \\ m^\ast & m^{\ast\prime} & m^{\prime\prime}
\end{pmatrix} = \delta_{mm^\ast}\delta_{m^\prime m^{\ast\prime}} \, ,
\end{equation}
\begin{multline}
\sum_{m m^{\prime}} \begin{pmatrix}
\ell & \ell^{\prime} & \ell^{\prime\prime} \\ m & m^{\prime} & m^{\prime\prime}
\end{pmatrix}
\begin{pmatrix}
\ell & \ell^{\prime} & \ell^{\ast\prime\prime} \\ m & m^{\prime} & m^{\ast\prime\prime}
\end{pmatrix} = \delta_{\ell^{\prime\prime}\ell{^\ast\prime\prime}}\delta_{m^{\prime\prime} m^{\ast\prime\prime}}  \\ \delta(\ell, \ell^\prime, \ell^{\prime\prime})\frac{1}{2\ell^{\prime\prime}+1} \, ,
\end{multline}
where, 
\begin{eqnarray}
\delta(\ell, \ell^\prime, \ell^{\prime\prime}) &=& 1  \,\,\,\,\, \textup{when} \,\, \vert \ell - \ell^\prime \vert \leq \ell^{\prime\prime} \leq \ell + \ell^{\prime} \, , \\
\delta(\ell, \ell^\prime, \ell^{\prime\prime}) &=& 0  \,\,\,\,\, \textup{otherwise}\, ,
\end{eqnarray}
substitute the kernel expansion Eqn.~\ref{kernelexp} and Eqn.~\ref{partalm} in Eqn.~\ref{cldef} we get 
\begin{equation}
\tilde{C}_{\ell} = \frac{2\ell^\prime + 1}{4\pi} \sum_{\ell^{\prime\prime}} (2\ell^{\prime\prime} + 1) C_{\ell^{\prime\prime}}^{M} \begin{pmatrix}
\ell & \ell^\prime & \ell^{\prime\prime} \\ 0 & 0 & 0
\end{pmatrix}^{2} C_{\ell^{\prime}} \,  , 
\label{partcl}
\end{equation}
where, $C_\ell^M$ is the smoothed mask ({\bf M}) power spectrum. Defining the mode-mode coupling 
matrix corresponding to the mask as,
\begin{equation}
M_{\ell\ell^\prime} = \frac{2\ell^\prime + 1}{4\pi} \sum_{\ell^{\prime\prime}} (2\ell^{\prime\prime} + 1) C_{\ell^{\prime\prime}}^{M} \begin{pmatrix}
\ell & \ell^\prime & \ell^{\prime\prime} \\ 0 & 0 & 0
\end{pmatrix}^{2}\, ,
\end{equation}
we finally arrive at
\begin{equation}
\tilde{C}_{\ell} = M_{\ell\ell^\prime}C_{\ell^{\prime}} \,  . 
\label{partcl}
\end{equation}
The importance of using Eqn.~\ref{partcl} is that we can estimate angular power spectrum corresponding 
to the unmasked region by multiplying the full-sky power spectrum with the mode-mode coupling matrix. 
Similarly, the converse is also true i.e.,
\begin{equation}
{C}_{\ell} = M_{\ell\ell^\prime}^{-1}\tilde{C}_{\ell^{\prime}}  \,  . 
\label{reccl}
\end{equation}
We can use Eqn.~\ref{reccl} to estimate the full-sky angular power spectrum from the partial-sky 
angular power spectrum given the mode-mode coupling matrix. 

During { the implementation of the Gibbs ILC method}, at any given Gibbs step $i$, we use Eqn.~\ref{reccl} to estimate the full-sky 
CMB angular power spectrum from the partial-sky CMB map. A new full-sky theoretical CMB angular power 
spectrum is then sampled from its conditional density as discussed in Section~\ref{GibbsILC} 
using the full-sky CMB angular power spectrum. Afterwards, we 
employ Eqn.~\ref{partcl} to estimate the partial-sky CMB angular power spectrum from the full-sky sampled
theoretical CMB angular power spectrum. This partial-sky CMB angular power spectrum is then 
used to sample a partial-sky cleaned CMB map. The details of sampling a partial-sky CMB map is presented 
in the following sub-section.

\subsubsection{Sampling a CMB map in a Partial-sky Analysis}
\begin{figure*}[!t]
\centering
\includegraphics[scale=0.6]{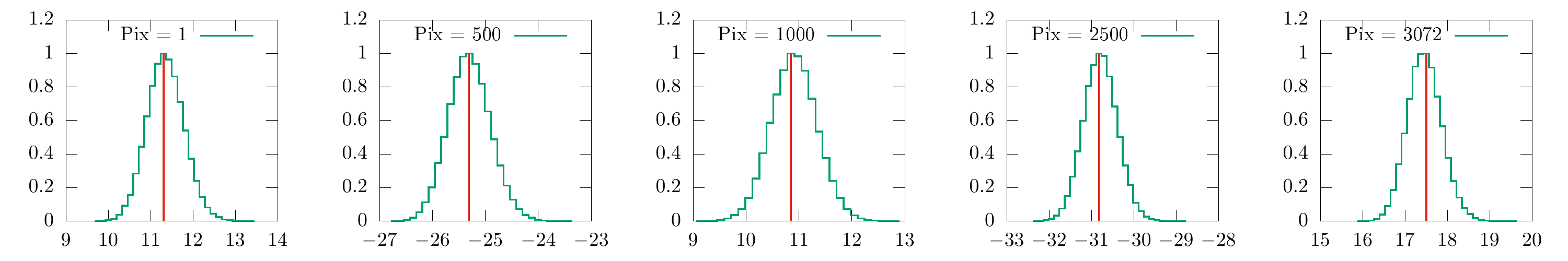}
\caption{The normalized probability density of CMB pixel temperatures for some selected pixels after the application of {\tt 
ThDust5000$\_$Eff} mask are shown in green.  The normalization is performed 
by dividing by the corresponding modal value such that the position of the 
peak corresponds to unity in all the plots. The horizontal axes represent pixel temperatures in the unit of $\mu$K (thermodynamic).  
The positions of mean temperatures are shown by the red vertical lines.}
\label{fig_data_norm_pix}
\end{figure*}

In~\cite{Saha2017}, we discussed how to implement global ILC method in an iterative manner in pixel 
space. Given that during a partial-sky analysis, the observed CMB maps at any given N$_{side}$ consist of 
$\tilde{\textup{N}}_{\tt pix}\, (\textup{where}, \,\tilde{\textup{N}}_{\tt pix} < {\textup{N}}_{\tt pix}$~\footnote{where $\textup{N}_{\tt pix}$ is the total number of pixels in a HEALPIX full-sky 
map at a given pixel resoultion $\textup{N}_{\tt side}$. The $\textup{N}_{\tt pix}$ and ${N}_{\tt side}$ is related as follows: 
N$_{\tt pix} = 12\times \textup{N}_{\tt side}^2$}) number of pixels, 
the theoretical CMB covariance weighted variance in this case in pixel space is as follows:
\begin{equation}
\tilde{\sigma}^2_{\tt GILC} = \tilde{\bf S}^{T} \tilde{\bf C}^{\dagger}\tilde{\bf S} \, .
\end{equation}
Here $\tilde{\bf S}$ is the a partial-sky cleaned CMB map with $\tilde{\textup N}_{\textup{pix}}$ of pixels. Similarly, 
$\tilde{\bf C}$ is a  ($\tilde{\textup{N}}_{\textup{pix}} \times \tilde{\textup{N}}_{\textup{pix}}$) matrix consisting of theoretical 
CMB covariance between only the surviving pixels. 

But in the Gibbs ILC algorithm using global ILC in pixel space is a computationally expensive procedure since 
at each Gibbs step we need to evaluate the theoretical CMB covariance matrix using the sampled 
theoretical CMB angular power spectrum. Hence, instead of using pixel space global ILC we implement 
the global ILC method in harmonic space (Eqn.~\ref{hilc1}) in the full-sky observations. 
For a partial-sky Gibbs ILC pipeline, we propose the following estimator based on the Eqn.~\ref{hilc_new} for estimating the weights:
\begin{equation}
\tilde{\sigma}_{r} = \sum_{i,j}w_i w_j \sum_{\ell=2}^{\ell_{\tt max}} (2\ell + 1) \frac{\tilde{\hat\sigma}_\ell^{ij}}{\tilde{C}_\ell^{\prime}} \, ,
\label{hilc_part}
\end{equation}
where $\tilde{\hat\sigma}_\ell^{ij}$ is the partial-sky cross-power spectrum corresponding to the $(i,j)^{\tt th}$ 
input CMB maps.

The partial-sky weights are calculated by minimizing Eqn.~\ref{hilc_part}. We use these new weights 
to estimate a partial-sky cleaned CMB map as follows, 
\begin{equation}
\tilde{\bf \hat{S} } = \sum_{i=1}^{n} w_i \tilde{\bf d}_i \, .
\label{cmap}
\end{equation}

\begin{figure*}[!t]
\centering
\includegraphics[scale=0.25]{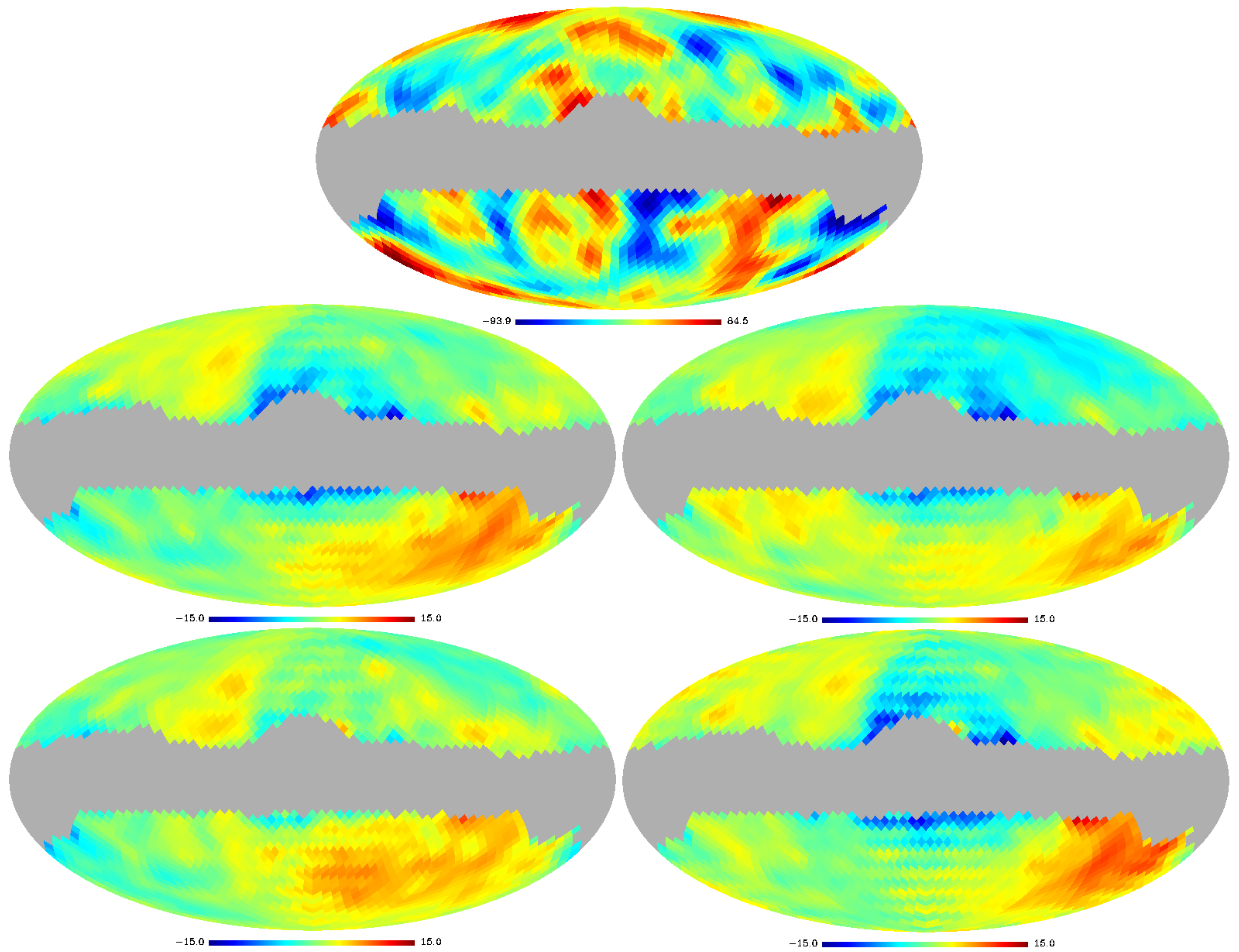}
\caption{The Best-Fit partial-sky cleaned CMB map obtained following our Gibbs ILC procedure at N$_{\textup{side}} =  16$  
and smoothed by a a Gaussian beam of FWHM = $9^\circ$ is shown in the top panel. While estimating the final best-fit map 
we masked the galactic region using our {\tt ThDust5000$\_$Eff} mask. In the middle left and middle right panel 
we show the difference maps comparing our best-fit map with Planck COMMANDER and NILC cleaned maps respectively. In the bottom panel 
from left to right we compare our best-fit map with SMICA and SEVEM cleaned maps respectively. All the difference maps are masked using  
{\tt ThDust5000$\_$Eff} mask.  All the  color scales are in $\mu$K thermodynamic temperature unit.}
\label{fig_all_data_maps}
\end{figure*}
\begin{figure}[!t]
\centering
\includegraphics[scale=0.25]{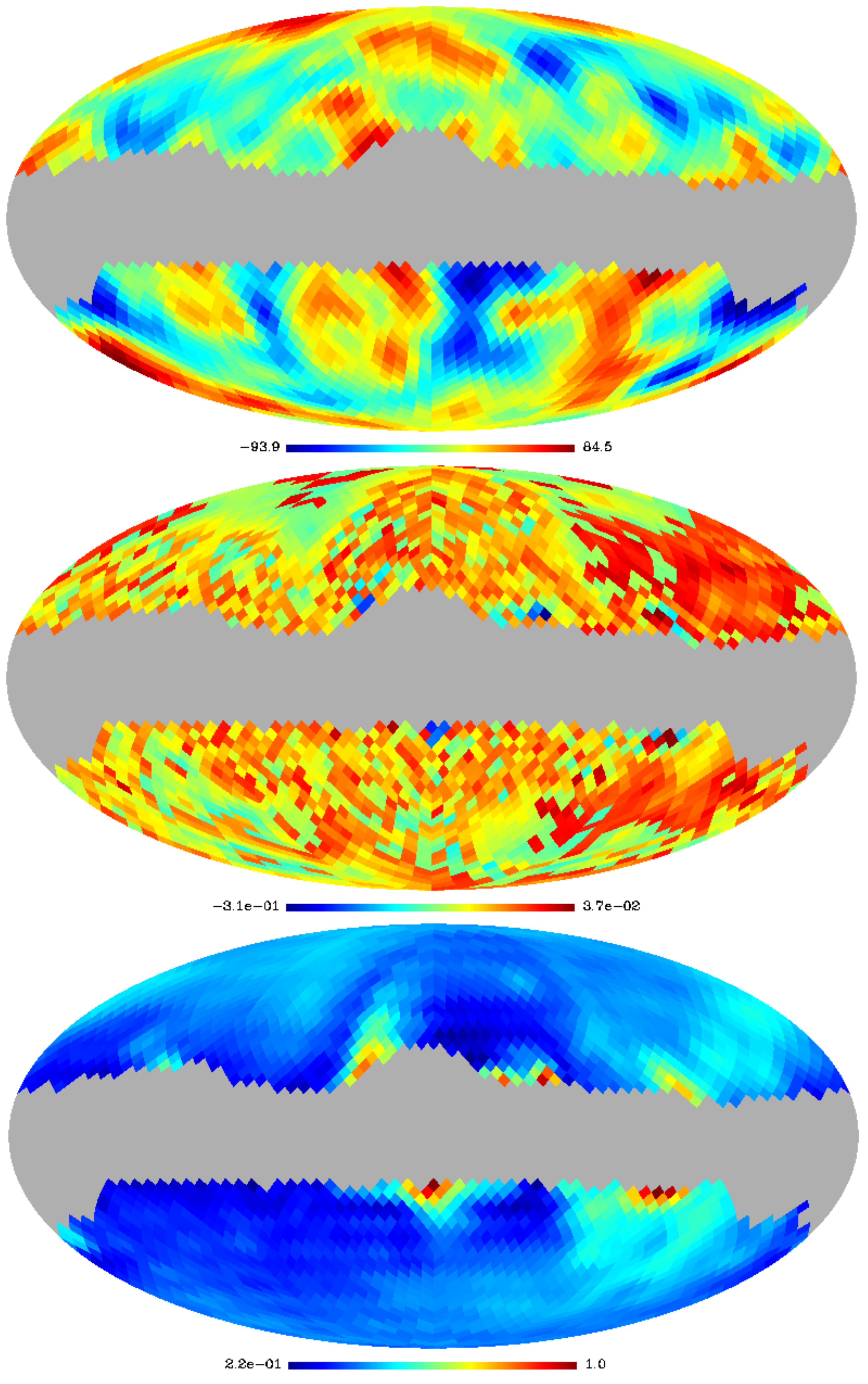}
\caption{Top panel shows the mean map estimated by taking the mean of all the partial-sky cleaned CMB maps  from our Gibbs ILC method. 
The  mean  map is then multiplied with {\tt ThDust5000$\_$Eff} mask. The mean map matches very well with the best-fit map shown in 
the top panel of Fig.~\ref{fig_all_data_maps} as shown in the the middle panel. The bottom  panel  shows  the  standard  deviation  
map  obtained  by  using  all  the cleaned maps masked by {\tt ThDust5000$\_$Eff} mask. All the  color scales are in $\mu$K thermodynamic temperature unit.}
\label{fig_data_mean_map}
\end{figure}
\begin{figure*}[!t]
\centering
\includegraphics[scale=0.6]{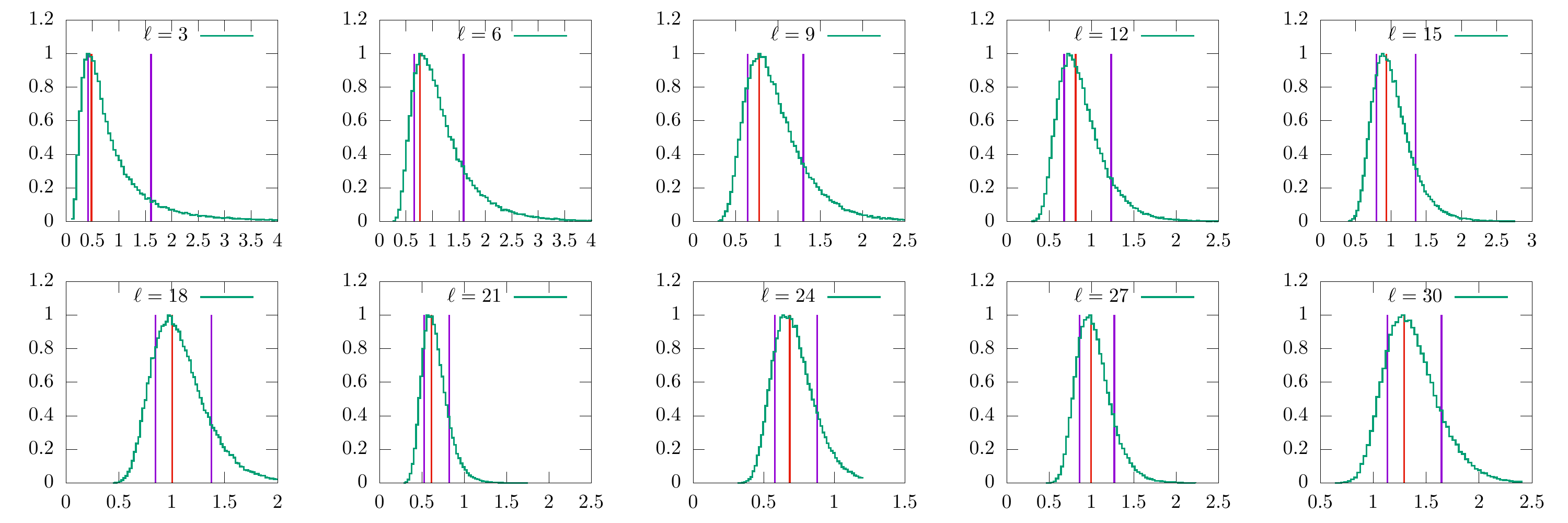}
\caption{We show the normalized densities of the CMB theoretical angular power spectrum obtained by Gibbs sampling 
for the mulitpoles corresponding to the bin-middle values. 
The normalization for each density is
such that the peak corresponds to a value of unity. 
The horizontal axis for each sub plot represents 
$\ell(\ell+1)C_\ell/(2\pi)$ in the unit 1000 $\mu$K$^2$.  The red line in each sub-plot corresponds to the mode-value of the 
corresponding probability density functions. The 1$\sigma$ confidence interval is marked by the region within the 
two vertical lines for the theoretical angular power spectrum.}
\label{fig_data_norm_l}
\end{figure*}
\begin{figure}[!t]
\centering
\includegraphics[scale=0.6]{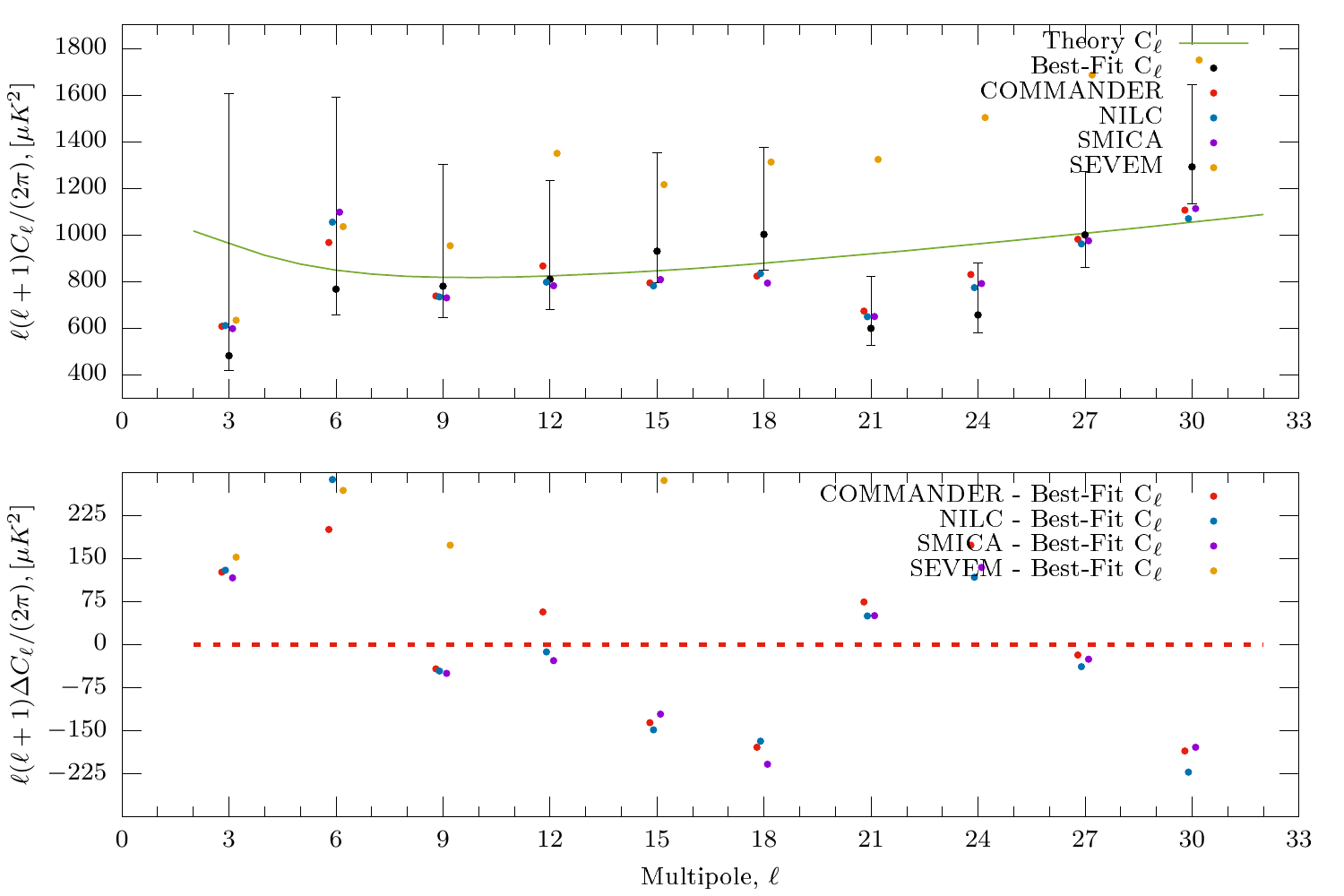}
\caption{In the top panel we show the binned best-fit CMB theoretical angular power spectrum along with the asymmetric error bars indicating 
68.27$\%$ confidence intervals obtained from the Gibbs samples in black line.  The binned angular power spectrum estimated from COMMANDER, NILC, 
SMICA and SEVEM cleaned CMB maps are shown with red, blue, violet and yellow points respectively.  (For visual purpose, these spectra are 
shifted along the horizontal axis slightly from their actual positions of the integer multipoles.)  The light green line shows the 
Planck-2018 theoretical angular power spectrum.  In the bottom panel, we show the differences of Commander, NILC, SMICA and SEVEM 
angular power spectra respectively from the best-fit angular power spectrum of top panel.}
\label{fig_data_cl}
\end{figure}

\section{Input Data}
\label{input}
\subsection{Input Maps}
In our analysis, we use all of the  WMAP nine-year difference assembly (DA) maps ~\citep{2011ApJS..192...14J} and seven Planck-2015 
maps - three of them correspond to  LFI frequencies ($30$, $40$ and $70$ GHz)~\citep{Ade:2015sua} and the rest 
are four HFI frequencies ($100, 143, 217$
and $353$ GHz)~\citep{Adam:2015vua}. The processing of all the input maps remains identical to~\cite{Saha2017} and 
results in a total of $12$ input maps (five at WMAP and seven at Planck frequencies).  All our input maps 
are at a pixel resolution N$_{\tt side}$ = $16$ and beam smoothed by a Gaussian beam  
of FWHM  = $9^\circ$ after properly taking taking care of the native beam resolutions. At this pixel and beam 
resolution we can ignore the 
contributions of noise. The monopole is removed from all the full-sky 
input maps before the analysis. 

\subsection{Masks}
\label{ourmask}
In the current analysis we use three different types of mask. 
The details of the various masks are given below. 

\subsubsection{ThDust5000}

 In order to  generate a mask based on the thermal dust emissions we follow the below steps.
 \begin{itemize}
 \item[1.] We downgrade Planck 70 and 353 GHz maps to N$_{\tt side} = 256$ and both at same effective beam resolution of 
FWHM = $6^{\circ}$.
\item[2.] We take the difference between Planck 353 and 70 GHz maps, which will give a map dominated by the thermal dust emissions.
\item[3.] We set all the pixels in the difference map with pixel value $\leq 5000~\mu$K to 1. All 
the remaining pixels are set to 0. 
 \end{itemize}
We refer to the mask obtained by following the above procedure as ``{\tt ThDust5000}'' and 
is shown in the left panel of Fig.~\ref{fig_ourmask}. Out of a total of 3072 pixels, the  {\tt ThDust5000} mask consists of 679 pixels 
with pixel value `0' and rest of the pixels are `1'.

\subsubsection{ThDust5000$\_$Sm \& and ThDust5000$\_$Eff}

As mentioned in the Introduction, using a binary mask such as {\tt ThDust5000} in the Gibbs algorithm results in Gibbs 
phenomenon at jump discontinuities across the mask boundaries. In order to avoid this situation, one feasible solution  is 
to smooth the mask. In this work, we use the Gaussian smoothing kernel for smoothing the mask
\begin{equation}
B(\theta) = \frac{1}{\sqrt{2\pi\sigma^2}} \textup{exp}\Big(-\frac{\theta^2}{2\sigma^2}\Big)\, ,
\end{equation} 
where $\theta$ is a separation angle. The Gaussian smoothing kernel can be considered as a low-pass filter with window function 
\begin{equation}
B_\ell = {\textup{exp}}(-\ell^2 \sigma^2)\, .
\end{equation}
Here, $\sigma = {\textup{FWHM}} / \sqrt{8\textup{ln}(2)}$ and FWHM of the smoothing kernel is set to 9$^{\circ}$ so that in our Gibbs ILC method,  
where we perform  a spherical harmonic expansion up to $\ell = 32$, all the the multipoles $\ell > 32$ are sufficiently suppressed. 
The smoothed mask {\tt ThDust5000$\_$Sm} obtained after smoothing the {\tt ThDust5000} mask 
 is shown in the middle panel of Fig.~\ref{fig_ourmask}. The operation of smoothing the mask causes 
some changes in the pixel values nearby the region boundaries in the mask. 
Some of the 0 valued pixels from the original 
unsmoothed mask might take small positive 
values and similarly the pixels with initial pixel value 1 might no longer be 1 after the smoothing operation. As we move away from the 
region boundaries the pixel values will no longer be affected significantly by the smoothing process. 

While estimating the final cleaned map we need to consider a new mask since some of the `1' valued pixels in {\tt ThDust5000} are 
no longer `1' in {\tt ThDust5000$\_$Sm} mask. Hence CMB estimated in those pixels will have strength lower than the actual value. In order 
to avoid the under estimation of CMB, we construct a new mask ``{\tt ThDust5000$\_$Eff}'' by setting all the pixels in {\tt ThDust5000$\_$Sm }
mask whose pixel value is $\ge 0.95$ as 1 and rest as 0. The {\tt ThDust5000$\_$Eff} consists of 1003 pixels with pixel value 0 out of 
total 3072 pixels.  

Unlike the present case, where full-sky CMB maps are available from WMAP and Planck satellite missions, if we 
consider a CMB mission (e.g.,  a ground based observatory) where only part of the entire sky can be observed, 
then the smoothing of the mask should be modified as follows. 
\begin{itemize}
\item[1.] Let Mask1 selects the portion of the sky observed by the experiment. Use Mask1 to construct a new mask Mask2 such that 
the new mask lies well inside of Mask1.
\item[2.] The area of the new mask, Mask2, should be selected in such a way that after Gaussian smoothing the smoothed Mask2 
pixel values in the unobserved region is = 0.
\end{itemize}

\section{Methodology}
\label{methodology}
We multiply all the input maps at a pixel resolution N$_{\tt side} = 16$ and smoothed by a Gaussian beam 
of FWHM = $9^{\circ}$ with the {\tt ThDust5000$\_$Sm}  mask (discussed in Section~\ref{ourmask}) thus mimicking 
partial-sky observed maps.  Our Gibbs ILC method 
has ten different chains, each with 10000 Gibbs steps.
The initial choice of $C_{\ell}$ for each  chain is made by  drawing  them uniformly within $\pm 3\Delta C_{\ell}$
around the Planck best-fit theoretical power spectrum~\citep{Aghanim:2019ame}, where $\Delta C_{\ell}$ 
denotes error due to cosmic 
variance alone.  Since our input maps are all partial-sky maps
, we cannot 
use the above sampled $C_{\ell}$ directly for computing the weights. For each chain, we  multiply the respective initial ${C_\ell} $
with the mode-mode coupling ($M_{\ell\ell^\prime}$) matrix  to obtain the corresponding partial-sky $C_\ell$. 
The $M_{\ell\ell^\prime}$ is estimated corresponding to the {\tt ThDust5000$\_$Sm}  mask, which relates a full-sky 
CMB angular power spectrum of a given map to the corresponding partial-sky angular power 
spectrum estimated from the {\tt ThDust5000$\_$Sm}  masked map.  
The partial-sky angular power spectrum is then used to sample ${\bf \tilde{\hat{S}}}$ using Eqn.~\ref{cmap} where the weights 
are estimated by minimizing Eqn.~\ref{hilc_part}.  
After sampling ${\bf \tilde{\hat{S}}}$ we estimate the partial-sky cleaned CMB power spectrum $\tilde{\hat C}_{\ell}$. 
In order to sample a full-sky theoretical CMB angular power spectrum, we use Eqn.~\ref{reccl} to estimate the 
full-sky angular power spectrum 
${\hat C_{\ell}}$ using $\tilde{\hat C}_{\ell}$. { This full-sky power spectrum is then mulitplied with 
$\ell(\ell+1)/2\pi$, beam and pixel window functions  and later binned 
with a bin width of 3 mulitpoles. The bin-middle values are at $\ell = 3, 6, 9, 12, 15, 18, 21, 24, 27\, \textup{and}\, 30$. 
After binning we divide the binned full-sky power spectrum by $\ell(\ell+1)/2\pi$, 
beam and pixel window functions. Now using this binned power spectrum we sample a new theoretical CMB angular power 
spectrum 
 by following Eqn.~\ref{cl_sample}.} The new sampled full-sky theoretical 
CMB angular power spectrum 
is then multiplied by the   $M_{\ell\ell^\prime}$ to obtain a sample of partial-sky power spectrum which 
is then used in the next Gibbs step.

At the end of all $10$ chains, we are left with a total of $100000$ joint samples of cleaned map and theoretical CMB 
power spectrum to sample the posterior density $P({\bf S}, C_{\ell}|{\bf D})$. The burn in phase 
in each chain is very brief. Visually, this phase does not 
appear to contain more than a few  Gibbs iterations. We, however, remove the initial $50$  
Gibbs iterations from each chain as a conservative estimate of burn-in period. After the burn-in
rejection, we end up with  a total of $99500$ samples from all chains.

\section{Results}
\label{Results}
\subsection{Cleaned Maps}
\label{cmaps}

In order to generate the best-fit cleaned CMB map we follow the procedure as outlined in~\cite{Sudevan:2018qyj}. We use all 
samples of cleaned CMB maps after rejecting those generated during the initial burn-in phase from each chain. We estimate the 
marginalized  probability density functions of CMB temperature corresponding to the pixels not masked out by the 
{\tt ThDust5000$\_$Eff} mask. 
We show the normalized probability density functions corresponding to some random pixels in Fig.~\ref{fig_data_norm_pix}. 
The normalization for each probability density function is such that the peak corresponds to a value of unity.
Taking the mode-value of the density function corresponding to each surviving pixel we form the best-fit partial-sky 
cleaned CMB map which  is shown in  the top panel of Fig.~\ref{fig_all_data_maps}. We compare  our 
best-fit cleaned CMB map with  
Planck Commander, NILC (needlet space ILC), SMICA and SEVEM 
cleaned maps~\citep{Akrami:2018mcd}. We downgrade the Planck cleaned maps to N$_{\tt side} = 16$ and smooth 
with a Gaussian beam of FWHM = $9^\circ$  after properly 
taking care of their  respective individual beams. 
We show the difference maps obtained after taking the difference between our best-fit cleaned map and 
the cleaned maps provided by Planck COMMANDER, NILC, SMICA and SEVEM 
in the  middle and bottom panels of the Fig.~\ref{fig_all_data_maps}. We remove the central 
galactic region by multiplying the maps with the {\tt ThDust5000$\_$Eff} mask. Clearly, from the difference map plots we 
can see that our best-fit map 
matches with rest of  the cleaned CMB maps quite well. 

Using all the 99500 partial-sky cleaned CMB maps from our Gibbs ILC method, we estimate a mean cleaned 
CMB map, the mean map. We show the mean map after multiplying with the {\tt ThDust5000$\_$Eff} mask 
in the top panel of Fig.~\ref{fig_data_mean_map}.
In the middle panel of Fig.~\ref{fig_data_mean_map} we show the difference between best-fit CMB map and mean CMB map.
We see that the mean map matches very well with the best-fit CMB map within an absolute difference of $< 1\mu K$. In order to quantify the 
reconstruction error while estimating a partial-sky cleaned CMB using Gibbs ILC method given partial-sky input CMB data 
and theoretical CMB angular power spectrum,
we generate a standard deviation map using all $99500$ partial-sky cleaned maps. We show this map in the bottom 
panel of Fig.~\ref{fig_data_mean_map}.  We can infer from this figure that the reconstruction error is very small all over the 
region of the sky  selected by {\tt ThDust5000$\_$Eff} mask. 
\begin{figure*}[!t]
\centering
\includegraphics[scale=0.6]{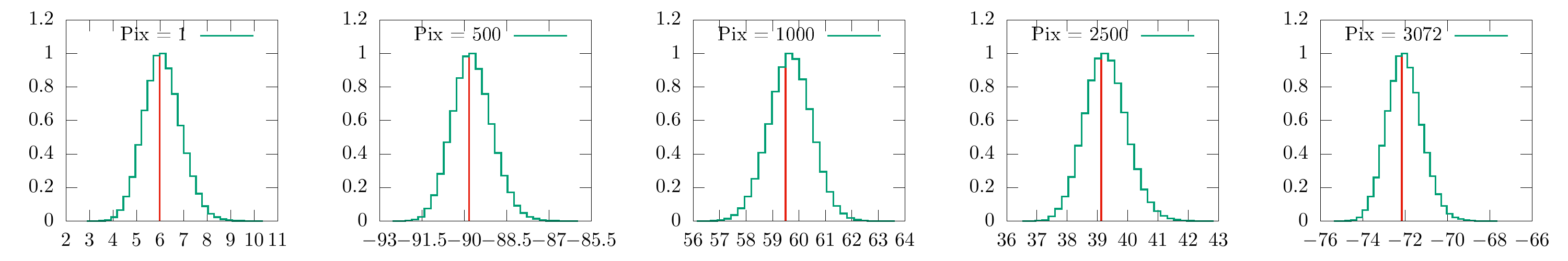}
\caption{The peak-normalized probability density of CMB pixel temperatures for some pixels survived after the application of {\tt 
ThDust5000$\_$Eff} mask are shown in green.  The horizontal axes represent pixel temperatures in the unit of $\mu$K (thermodynamic).  
The positions of mean temperatures are shown by the red vertical lines.}
\label{fig_sim_norm_pix}
\end{figure*}
\begin{figure*}[!t]
\centering
\includegraphics[scale=0.2]{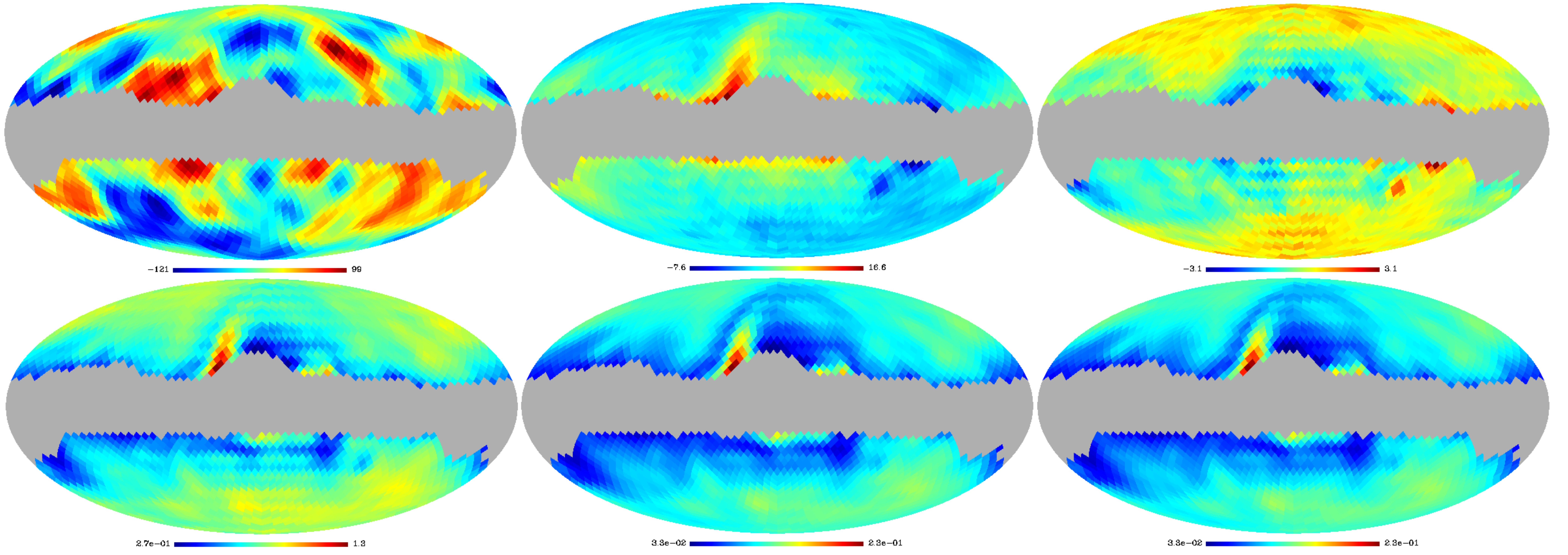}
\caption{In the top left panel we show the best-fit cleaned CMB map obtained from the first Monte Carlo simulation, 
followed by the difference map obtained after taking the difference between the best-fit cleaned CMB map the input CMB map used 
in that particular simulation. In the top right panel we show mean of all 1000 such difference maps estimated from each Monte Carlo 
simulation.  In the bottom right pane, we show the standard deviation map from the first simulation. The standard deviation map 
signifies the error while reconstructing a cleaned CMB map following our Gibbs ILC algorithm. In the bottom middle and bottom right panels 
we show the mean and standard deviation of all such 1000 standard deviation maps. All the maps that we show in this figure are 
multiplied by  {\tt ThDust5000$\_$Eff} mask. All the  color scales are in $\mu$K thermodynamic temperature unit.}
\label{fig_sim_maps}
\end{figure*}
\subsection{Angular Power Spectrum}
\label{theorycl}

We estimate the marginalized probability density of CMB theoretical angular power spectrum  
 for multipoles corresponding to the bin-middle values and show the respective density functions in Fig.~\ref{fig_data_norm_l}. 
We normalize these density functions to a value of unity at their peaks. 
The horizontal axis of each plot of this figure represents $\ell\left(\ell +1\right)C_{\ell}/(2\pi)$ in the unit 
of $1000$ $\mu K^2$. The density functions show long asymmetric tails for low multipoles (e.g., $\ell = 3, 6, 9$). For large
 multipoles ($\ell \ge 21$) the asymmetry of the densities become gradually reduced. The region within the two vertical lines in each plot 
 corresponds to the $1\sigma$ ($68.27\%$) confidence interval for the CMB theoretical power spectrum for each multipoles.      

In the top panel of Fig.~\ref{fig_data_cl} we show the binned best-fit theoretical CMB angular power spectrum, with black colored points, 
defined by the positions of peaks of marginalized angular power spectrum density functions. The asymmetric error bars at each 
binned $\ell$ values 
show the $1\sigma$ confidence interval for the theoretical angular power spectrum. The light green line 
shows the theoretical power spectrum consistent with Planck 2018 results~\citep{Aghanim:2019ame}.
The best-fit theoretical angular power spectrum agrees well with the binned spectra estimated from Commander, 
NILC and SMICA cleaned maps, which are shown by red, blue and violet points respectively. 
In the bottom 
panel of  Fig.~\ref{fig_data_cl},  we show the difference of the best fit and Commander NILC, SMICA and SEVEM angular power spectra. 
We emphasize here that the all the Planck cleaned maps angular power spectra are estimated from the respective full-sky maps.



\section{Monte Carlo Simulations}

\label{MC}

For our detailed Monte Carlo simulations, we generate 1000 sets of input maps at the $12$ different WMAP and Planck frequencies at 
a Gaussian beam resolution $9^{\circ}$ and N$_{\textup side} = 16$ 
following the procedure as described in~\cite{Saha2017} and~\cite{Sudevan:2018qyj}. Our foreground model 
consists of synchrotron, free-free and thermal dust emissions. We use Planck 2018 foreground 
templates for generating foreground maps at all WMAP and Planck frequencies. The
random CMB realizations that we use 
in the input frequency maps are generated using the CMB theoretical angular power spectrum 
consistent with Planck 2018 results.  Once we simulate the foreground contaminated CMB maps, we then 
proceed to mask all the 1000 sets of 12 input maps 
using the {\tt ThDust5000$\_$Sm} mask. The masked input maps are then used in our Gibbs ILC code which consists of     
a total of $10$ Gibbs chains, each with 10000 Gibbs steps.  We follow the same methodology as adopted in the 
real data.


As was the case with real data, the initial burn-in period ends quickly. 
We continue keeping a conservative limit of initial $50$
samples corresponding to the burn-in phase which are then removed from each Gibbs chain. After 
initial burn-in rejection, we are left with a total of $9950$ joint samples of
cleaned map and theoretical angular power spectrum from each Gibbs chain.

\begin{figure*}[!t]
\centering
\includegraphics[scale=0.6]{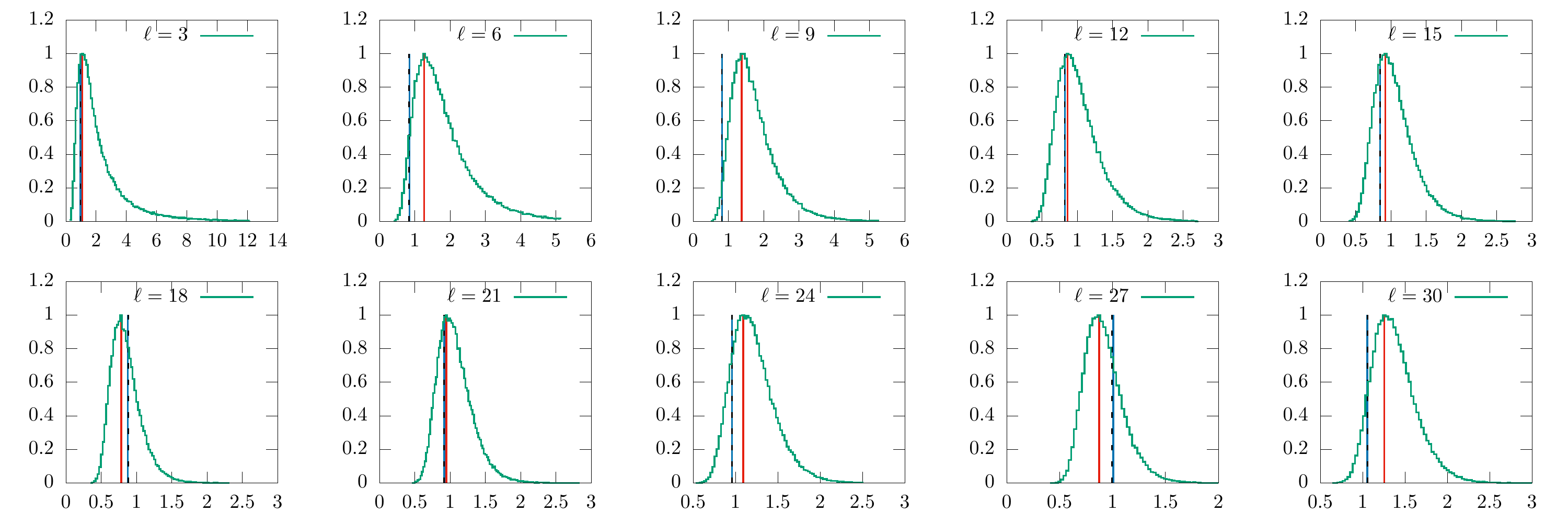}
\caption{The peak-normalized densities of the CMB theoretical angular power spectrum estimated from first Monte Carlo simulation 
for the mulitpoles corresponding to the bin-middle values. The horizontal axis for each sub plot represents 
$\ell(\ell+1)C_\ell/(2\pi)$ in the unit 1000 $\mu$K$^2$.  The red line in each sub-plot corresponds to the mode-value of the 
corresponding probability density functions. 
The blue line represents the mean of all 1000 mode values corresponding to each density function and the black dashed line is 
the value of Planck-2018 theoretical CMB angular power spectrum.}
\label{fig_sim_norm_l}
\end{figure*}
\begin{figure}[!t]
\centering
\includegraphics[scale=0.5]{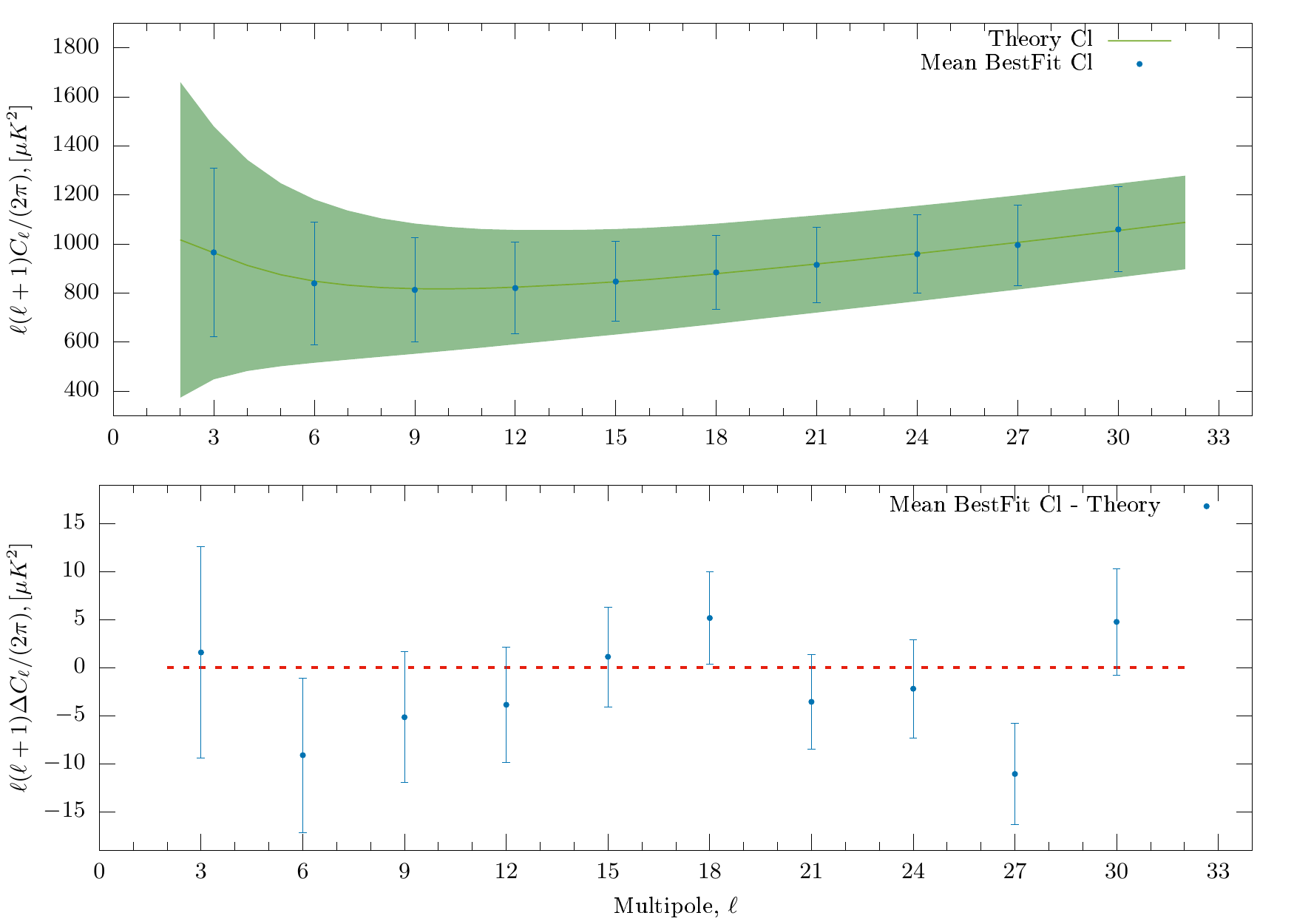}
\caption{In the top panel we show the mean of all 1000 best-fit estimate of theoretical CMB angular power spectrum  with blue points.  
The green line indicates the Planck-2018 theoretical CMB angular power spectrum. We show the difference  between the mean best-fit 
angular power spectrum an Planck-2018 theoretical angular power spectrum in bottom panel.}
\label{fig_sim_cl}
\end{figure}

Using all cleaned map samples from all chains after burn-in rejection we form marginalized density 
functions corresponding to each surviving pixels in a $N_{side} = 16$ map.  We show the normalized density 
functions corresponding to some random  
pixels from a representative simulation (here for instance, the Monte Carlo Simulation No. 1) 
 in Fig.\ref{fig_sim_norm_pix}. A CMB map formed from the pixel 
temperatures corresponding to the modes of these density functions define the best-fit 
partial-sky CMB cleaned map obtained from the simulation. We only consider those pixels which have pixel value 1 in 
the {\tt ThDust5000$\_$Eff} while estimating the best-fit cleaned map. We show our partial-sky best-fit 
cleaned map from the above simulation in the top-left panel 
of Fig.~\ref{fig_sim_maps}. The difference of the best-fit and input CMB realization
after masking with the {\tt ThDust5000$\_$Eff} mask is shown in the top middle panel of the 
same figure. Clearly, the best-fit CMB map matches very well
with the input CMB realization used in the simulation. The maximum difference ($\sim 10$ $\mu K$) 
between the two maps is observed along the edges of the {\tt ThDust5000$\_$Eff} mask. 
This shows that our method
removes foreground reliably in the { outer regions of the Galactic plane}. 
The top right panel of Fig.~\ref{fig_sim_maps} shows the mean of all the 1000 difference maps 
estimated by taking the mean of difference between best-fit and input CMB maps 
obtained from all Monte Carlo simulations.  
The bottom right panel of Fig.~\ref{fig_sim_maps} shows the  error map
computed from the Gibbs samples. The error map is a standard devaition 
map computed using all the cleand CMB samples from Monte Carlo simulation No. 1. 
The maximum error of $1$ $\mu K$ is observed along the 
edges of the {\tt ThDust5000$\_$Eff} masked region. We show the mean and standard 
deviation of 1000 such error maps from each Monte Carlo simulations in the bottom middle 
 and bottom left panels of Fig.~\ref{fig_sim_maps}. We see that the error while reconstructing  
  a cleaned CMB pixel following our partial-sky Gibbs ILC method is very small. 
In summary, using the Monte-Carlo simulations, we see that our Gibbs ILC method reliably minimizes the  foreground 
across all the 1000 simulations.

We show normalized density functions corresponding to 
the bin-middle multipoles from some 
representative simulation in Fig.\ref{fig_sim_norm_l}. 
The normalization is performed such that the peak value of the density functions are set to unity. 
The horizontal axis for each sub-plot represents 
$\ell(\ell+1)C_\ell/(2\pi)$ in the unit 1000 $\mu$K$^2$.  The red line in each sub-plot corresponds to the mode-value of the 
corresponding probability density functions. 
The blue line represents the mean of all 1000 mode values corresponding to each density function and the black dashed line is 
the value of Planck-2018 theoretical CMB angular power spectrum.  
We show the mean of the best-fit estimate of binned underlying CMB theoretical angular power 
spectrum from all  the 1000 Monte Carlo simulations in Fig.~\ref{fig_sim_cl} in 
blue points. The multipole values corresponding to each point is the bin-middle value. 
 The underlying theoretical angular power spectrum 
is shown in light green. The blue vertical lines shows the standard deviation corresponding 
to each multipoles.   

\section{Conclusions \& Discussions}
\label{Conclusion}
In this article, we formalize a foreground model independent approach
to estimate the posterior density of CMB map and the corresponding
theoretical angular power spectrum following Gibbs-ILC method using
sky region outside the galactic plane. We use all the
10 detector set maps provided by WMAP and 7 frequency maps from
Planck mission (LFI channels - 30, 44, 70 GHz and HFI -
100, 143, 271, 353 GHz). Before foreground removal all the input maps are preprocessed
as discussed in Section 3 which results in a total of 12 input
maps at Healpix pixel resolution N$_{\textup{side}} = 16$ and smoothed by a Gaussian beam of
FWHM = 9$^\circ$.

We exclude the galactic region of the sky by multiplying the
input maps with a smoothed mask.  Using smoothed mask mitigates the adverse
effects due to improper behavior of spherical harmonic transforms at jump
discontinuities around the mask boundaries. There are two major outcomes
of generalizing the full-sky Gibbs ILC method over the partial-sky. First,
the weights that are the primary variables to remove the foregrounds, can
now be more effectively tuned to minimize the foregrounds based upon the
local information of foreground strength or  depending upon the nature of foreground contaminations
in input maps. Secondly, the method can now be applied on the CMB observations
which can produce maps only over a fraction of region of the entire sky (e.g.,
as expected from ground based CMB observations). Finally, the final CMB
products of this new work are independent of detailed foreground
modelling.

We perform detailed Monte Carlo simulations to validate the Gibbs ILC method
over partial-sky.  We simulate 1000 noise and foreground contaminated CMB maps
(based on the foreground model provided by Planck 2018 results) at the same WMAP and
Planck frequencies which  are included in the case of foreground removal from real data.
The simulation results show that our method effectively and accurately removes the
foregrounds present in the unmasked region. The mean of all binned best-fit theoretical
angular power spectra from all simulations agree excellently with the input Planck-2018
theoretical CMB angular power spectrum used to generate the random realizations of CMB
maps.

For the case of real data, using all the 99500 samples of partial-sky cleaned CMB
maps and binned estimates of full-sky CMB angular power spectra we estimate the
best-fit (partial-sky) cleaned map and best-fit theoretical CMB angular power spectrum
respectively. We see a nice agreement between our results and those obtained from the
Planck COMMANDER, NILC, SMICA cleaned CMB maps. 
 
This  work is based on observations obtained with Planck (http://www.esa.int/Planck)
and WMAP (https://map.gsfc.nasa.gov/). Planck is 
an ESA science mission with instruments and contributions directly funded by ESA Member States, 
NASA, and Canada.   
We acknowledge {the} use of Planck Legacy Archive (PLA) and the Legacy Archive for Microwave Background 
Data Analysis (LAMBDA). LAMBDA is a part of the High Energy Astrophysics Science Archive Center (HEASARC). 
HEASARC/LAMBDA is supported by the Astrophysics Science Division at the NASA Goddard Space Flight
Center. We  use publicly available HEALPix~\cite{Gorski2005} package  
(http://healpix.sourceforge.net) for the analysis of this work.


\end{document}